\documentclass[useAMS,usenatbib]{mn2e}
\usepackage[dvips]{graphicx}
\usepackage{epsfig}
\usepackage{amsmath}
\title[Identification of substructure in phase-space]
{On the identification of substructure in phase-space using orbital frequencies
} 

\author[F. A. G\'omez \& A. Helmi]{Facundo A. G\'omez\thanks{Email:gomez@astro.rug.nl} 
\& Amina Helmi \\
Kapteyn
Astronomical Institute, University of Groningen, P.O. Box 800, 9700 AV
Groningen, The Netherlands}

\begin{document}

\date{}

\pagerange{\pageref{firstpage}--\pageref{lastpage}} \pubyear{}

\maketitle

\label{firstpage}

\begin{abstract}
  We study the evolution of satellite debris to establish the most
  suitable space to identify past merger events. We confirm that the
  space of orbital frequencies is very promising in this respect. In
  frequency space individual streams can be easily identified, and
  their separation provides a direct measurement of the time of
  accretion. We are able to show for a few idealised gravitational
  potentials that these features are preserved also in systems that
  have evolved strongly in time. Furthermore, this time evolution is
  imprinted in the distribution of streams in frequency space. We have
  also tested the power of the orbital frequencies in a fully
  self-consistent (live) N-body simulation of the merger between a disk
  galaxy and a massive satellite. Even in this case streams can be
  easily identified and the time of accretion of the satellite can be
  accurately estimated.
\end{abstract} 


\section{Introduction}

In the current standard cosmological model, known as $\Lambda$CDM,
galaxies like our own Milky Way are formed bottom-up, through the
merger and accretion of smaller building blocks  that come
together due to their gravitational attraction \citep[e.g.][]{wr}.

This model is being continuously tested and many questions and puzzles
remain to be addressed. An example is the large number of bound
substructures predicted to orbit galaxies like the Milky Way and M31
compared to the observed abundance of satellite galaxies in these
systems \citep[][]{kly,moore}, although several solutions have been
proposed \citep[e.g.][]{bullock2000}.  The formation of realistic disk
galaxies in full cosmological simulations remains another great
challenge \citep[e.g.][]{sn} yet steady progress is being made
\citep[e.g.][]{gov04,gov07,scannapieco09}.

One strong test of the current paradigm may be performed through
observations of the stellar halos of galaxies like the Milky Way. For
example, if the Galaxy was formed in a hierarchical fashion via
mergers, then we should be able to find fossil signatures of these
events in the present day phase-space distribution of halo stars. This
idea was strongly exalted after the discovery of the disrupting Sgr
dwarf galaxy by \citet{ibata94}. In the following years further debris
from accretion events was discovered, not only in the halo
\citep{hwzz99,grill,belu} but also in the disk of our Galaxy
\citep{eggen,ibata03,yanni,nhf,h06}, although some of this substructure
may be of dynamical origin \citep{ant09,min09}.

However, if the hierarchical scenario is correct, we are still far
from having unveiled every single stream associated to each merger
event the Galaxy has experienced in its lifetime. \citet{h03} by
combining numerical simulations with analytic work predicted that
there should be several hundreds of cold stellar streams present in
the vicinity of the Sun \citep[see also][]{vol08}.  The short
dynamical time scales, especially in the Solar neighbourhood, are the
basic reason behind why such a large number is expected. An accreted
satellite, orbiting in the inner regions of the Galaxy will give rise
to multiple stellar streams in a relatively short period of time. As
shown by \citet{hw} the spatial density of these streams is a strongly
decreasing function of time. Consequently, substructure associated to
a past accretion event is expected to have a very low-density
contrast, making its detection very difficult from its distribution in
space. On the other hand, due to the conservation of phase-space
density, the velocity dispersion of a stream will decrease as time
goes by, making the streams tighter in the velocity domain.

The arguments above highlight the need for full 6D phase-space
information to completely disentangle the predicted wealth of
substructure. Furthermore, samples of at least 1,000 halo stars would
appear to be necessary for this enterprise. Statistical arguments
using the revised NLTT proper motion survey \citep[New Luyten Two
Tenths,][]{gs03,sg03} have been used to measure the amount of
granularity (associated to moving groups) in the stellar halo near
Sun. \citet{gould03} has shown that no streams are present in this
region of the Galaxy that contain more than approximately 5\% of the
halo stars near the Sun. This estimate is in good agreement with the
expectations described above, but does not provide a direct
confirmation of the predictions. Spectroscopic surveys such as RAVE
\citep{zwitter08} and SEGUE \citep{yanny09} should be helpful in
identifying further nearby streams \citep[e.g.][]{smith09,klement} but
the breakthrough will take place only with the astrometric satellite
{\it Gaia} \citep{gaia}. {\it Gaia}, expected to be launched in 2011,
will measure the positions and motions of $10^9$ stars with very high
accuracy.

Given the datasets that soon will become available it is natural to ask
what is the optimal way to discover merger debris. More precisely, the
question we would like to address in this work is what is the best
space to look for substructure if our goal is to disentangle how the
Galaxy was assembled.

The first steps in this direction have been presented by
\citet{hz00}. These authors analysed simulations of the disruption of
satellite galaxies in a fixed Galactic potential and concluded that
the ``integrals-of-motion'' space defined by the energy, the total
angular momentum and its $z$-component $(E,L,L_{z})$ is very suitable
for identifying the remains of past accretion events. Even in fully
self-consistent cosmological simulations, disrupted satellites still
appear as coherent structures in this ``integrals of motion'' space
\citep{knebe05}. \citet{font} demonstrated the power of complementary
chemical abundance information to disentangle debris from different
events. \citet{ari_fu} have used for samples of nearby stars the
space defined by $(U^2 + 2 V^2)^{1/2}$ and $V$, where $U$ is the
radial velocity in the Galactic plane, and $V$ is the velocity 
component in the direction of rotation. More recently \citet{mcm} 
(hereafter MB08) showed that action-angles coordinates, and in 
particular, the orbital frequencies could conform a very convenient 
set of variables to identify nearby substructure.

In this paper we explore further the power of the orbital
frequencies. We start our journey by briefly reviewing the concept of
action-angle variables in Section 2. In Section 3 we discuss in depth
the space of orbital frequencies.  We characterise the distribution of
debris in a few simple cases and in particular, in a static spherical
potential. In Section 4 we analyse how the structure of this space is
altered in a time dependent (still rigid) potential. Finally, in
Section 5 we focus on a live N-body simulation of the accretion of a
massive satellite onto a pre-existing thin disk. We discuss and
summarise our results in Section 6.

\section{The action-angle variables}
\label{sec:act-ang}
Let us consider a dynamical system with a time-independent 
Hamiltonian $H({\textbf{\textit p}},{\textbf{\textit q}})$, where $({\textbf{\textit p}},{\textbf{\textit q}})$ 
is a set of $2n$ canonical coordinates. The dynamical state of 
this system is governed by the Hamilton's equations
\begin{equation}
\label{hamilton}
\displaystyle
\dot{p_{i}}=-\frac{\partial H}{\partial q_{i}},
\qquad \dot{q_{i}}= \frac{\partial H}{\partial p_{i}},
\end{equation}
with $i=1,\dots,n$.

The evolution of a system may be expressed much more simply by 
performing a canonical transformation of coordinates such that 
the equations of motion are 
\begin{equation}
\label{hamilton_prim}
\displaystyle
\dot{P_{i}}=-\frac{\partial H({\bf{P}})}{\partial Q_{i}}=0, \qquad
\dot{Q_{i}}= \frac{\partial H({\bf{P}})}{\partial P_{i}}=\Omega_{i}({\bf{P}}),
\end{equation}
with solutions
\begin{equation}
\displaystyle
\label{time_evol}
Q_{i}=Q_{i}^{0}+\Omega_{i}t,\qquad
P_{i}=P_{i}^{0}.
\end{equation}
If the system under study is such that the motion is periodic, 
then this set of canonical variables is known as action-angle variables. 

\begin{figure*}
\vspace{0.1cm}
\centering
\includegraphics[width=170mm,clip]{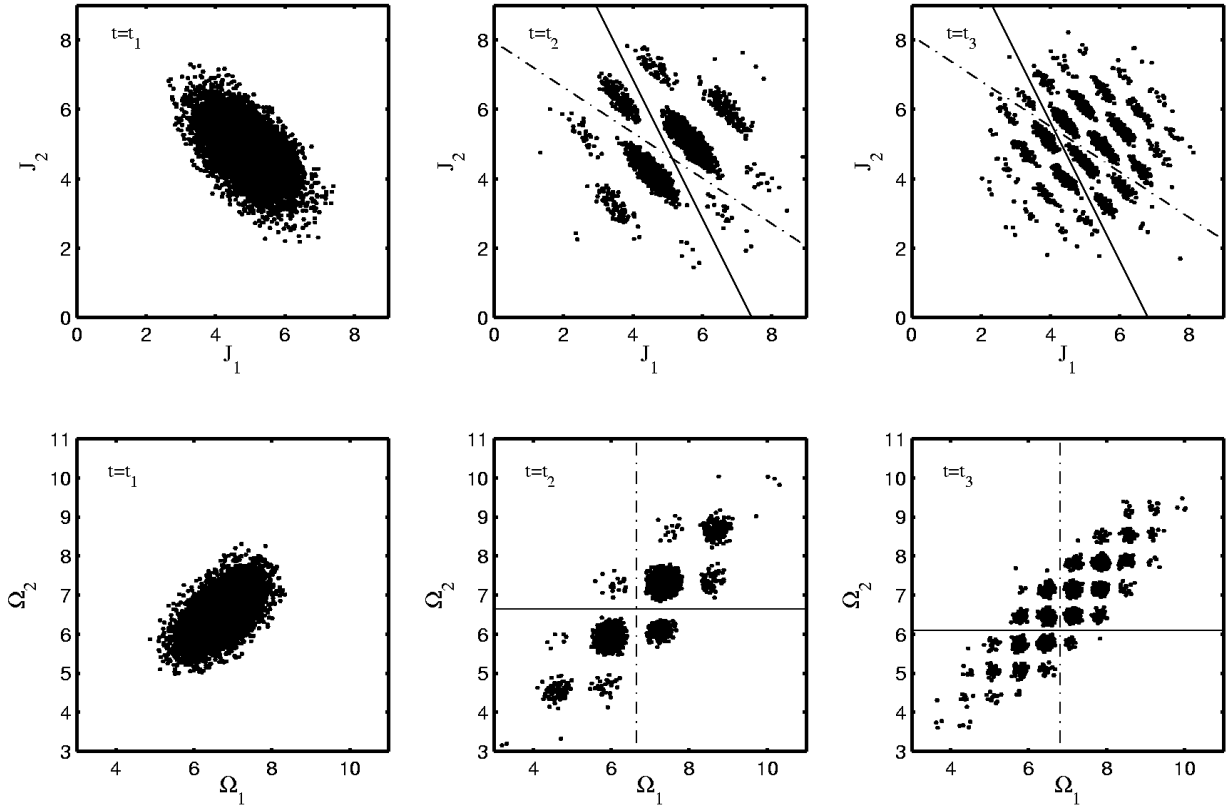}
\caption{Distribution of particles in actions (upper panels) and
  frequency (lower panels) space located in a small region of the
  angle space at three different times. Each clump represents a
  particular stream crossing this region at the time considered. The
  solid lines indicate $\Omega_{2} = cst$ whereas the dotted-dashed
  lines represent lines of constant $\Omega_{1}$, as shown in the
  bottom panels.}
\label{fig:acts_om_evol}
\end{figure*}

In a spherical potential $\Phi(r)$, the radial action is
\begin{equation}
\label{acts_rad}
\displaystyle
J_{r}=\frac{1}{\pi}\int_{r_{1}}^{r_2}{dr\frac{1}{r}\sqrt{2[E-\Phi(r)]r^{2}-L^{2}}}
\end{equation}
where $L$ is the total angular momentum, $r_{1}$ and $r_{2}$ the
turning points in the radial direction and
\begin{equation}
\displaystyle
\label{acts_ang}
J_{\phi}=L_{z},\qquad
J_{\theta}= L - L_{z}
\end{equation}
are the angular actions. The frequencies of motion $\Omega_{i}
= \partial H/\partial J_{i}$, can be derived by differentiation of the
implicit function
\begin{equation}
\displaystyle
g=J_{r}-\dfrac{1}{\pi}\int_{r_{1}}^{r_2}{dr\dfrac{1}{r}\sqrt{2[E-\Phi(r)]r^{2}-L^{2}}}
\end{equation} 
defined by Eq.~(\ref{acts_rad}). For more details, see
e.g., \citet{goldstein}.

\section{Frequency space for static potentials}

\subsection{Toy-Model}
\label{sec:toy_model}

The space of orbital frequencies in the context of isolating debris from 
accreted satellites was first introduced by MB08. For completeness we here
review its main characteristics. We also refer the reader to Section 4 of
MB08. For this purpose we have developed a simple toy-model and followed 
the evolution of a satellite in this space.

Let us consider a satellite (i.e. a system of particles that are
initially strongly clustered in phase-space) orbiting under the
influence of an arbitrary gravitational potential. For simplicity, we
consider a system with only two independent frequencies of motion. We
will also assume that the initial distribution in action-angle space
of the satellite follows a multivariate Gaussian with dispersions of
$\sigma_{\theta} = \pi/4$ and $\sigma_{J} = \pi/3$ in the angles and
actions, respectively (and no correlation among the different
directions).  In this space, the location of a given particle will
vary only due to the evolution of the angle coordinates. The rate of
evolution, i.e. the frequencies, is completely determined by the
underlying potential in which our satellite evolves (see
Eq.~(\ref{time_evol})). The other two coordinates, i.e., the actions,
will remain constant in time.

Here we shall assume that we can express the frequencies in the following way:
\begin{equation}
\displaystyle
\begin{array}{ll}
\label{in_freqs}
\Omega_{1} = c_1 J_{1} + c_2 J_{2},\\
\\
\displaystyle
\Omega_{2} = c_3 J_{1} + c_4 J_{2},
\end{array}
\end{equation}
where $c_i$ are constants. In general, the relation between the
frequencies and the actions is more complex, and depends on the
specific form of the gravitational potential. 

Typically, we have access to samples of stars located in a particular
region of space such as, for example, the Solar neighbourhood.  This
constraint acts as a filter, implying that we only observe the subset
of stars with orbital properties such that they are in this particular
location at the time of observation. To mimic this, we focus on the
particles located in a small region of the angles' space in our
toy-model. Figure~\ref{fig:acts_om_evol} shows the distribution of
these particles in the space of actions and of frequencies at three
different times ($0 < t_{1} < t_{2} < t_{3}$).

As discussed by MB08 the particles passing through a particular
location of (angle-)space are distributed in different patches or
streams. With time, the number of streams increases while their extent
in frequency space decreases. When the particles cross our location
for the first time, they have not had enough time to become
significantly spread in angle. We then observe only one big stream
(left panel of Fig.~\ref{fig:acts_om_evol}). At some later time,
particles with the highest frequencies begin to overtake the slowest
ones and therefore we observe multiple streams (since the angles are
periodic) each with its own characteristic frequency. The size of each
individual stream at any given time is dependent on the size of the
small region under consideration and on the initial spread of the system.

Since the orbital frequencies are constant (for a time independent
gravitational potential) the location of a particle in frequency space
remains unchanged.  The changing pattern shown in
Fig.~\ref{fig:acts_om_evol} is simply a consequence of the fact that
we restrict our analysis to particles at a given spatial location. The
gaps thus correspond to particles which are located outside our
window.

The network of patches is shaped by lines of constant frequency
$\Omega_{i} = cst$. As a consequence, and because of the relation
implied by Eq.~(\ref{in_freqs}), the distribution of particles in
action space is also very regular. However, in general the relation
between frequencies and actions will not be linear, and therefore the
distribution of particles in action space is typically more complex
(see e.g. Figure~\ref{fig:comp} below).

The number of clumps observed along each frequency direction at a
given location can be estimated by considering the initial spread in
the frequency. For example, at $\Omega_{2} = cst$, $\Delta \Omega_{1}
= \Omega_{1}^{\rm max} - \Omega_{1}^{\rm min}$.  After some time $t=t_{3}$,
the angle $\theta_{1}$ between the particles with the largest and
smallest $\Omega_{1}$ has increased to
\begin{equation}
\displaystyle
\Delta \theta_{1}(t_{3}) = \Delta \Omega_{1} t_{3} + \Delta \theta_{1}(t_{0}).
\end{equation}
with $\Delta \theta_{1}(t_{0})$ the initial separation at time $t_{0}$. Since 
$\theta_{1}$ is a $2\pi$-periodic variable the number of streams $N_{1}$ at 
$\Omega_{2} = cst$ is
\begin{equation}
\displaystyle
N_{1} = \left[\frac{\Delta \theta_{1}(t_{3})}{2\pi}\right] = \left[\frac{\Delta \Omega_{1} t_{3} +  \Delta \theta_{1}(t_{0})}{2\pi}\right]  
\approx \left[\frac{\Delta \Omega_{1} t_{3}}{2\pi}\right]
\end{equation}
where the last approximation holds if $t_{3}$ is long enough such that
$\Delta \Omega_{1} t_{3} >> \Delta \theta_{1}(t_{0})$.

Another quantity of interest is the distance between two adjacent streams in 
frequency space, 
\begin{equation}
\displaystyle
\delta \Omega_{i} = \frac{\Delta \Omega_{i}}{N_{i}} \approx \frac{2\pi}{t}, \qquad i=1,2.
\label{sep}
\end{equation}
The time $t$ may be considered as the time since disruption of our system
(i.e. its constituent stars have been tidally stripped, and move only
under the influence of the host gravitational field). Eq.~(\ref{sep})
states that this time can be directly estimated by measuring $\delta
\Omega_{i}$. Moreover, systems will have different values of $\delta
\Omega_{i}$ depending on the time since disruption, which implies that
this characteristic scale could be used to recognise the streams
originating in objects plausibly accreted at different epochs.

\subsection{Spherical potentials}

We now consider dynamical systems with an underlying
spherical potential, $\Phi(r)$. Due to the conservation of angular
momentum $L$, the motion of stars are constrained to remain in a
plane, implying that there are only two independent frequencies of
motion: $\Omega_{r}$ associated to the radial oscillation and
$\Omega_{\phi}$ associated to the angular oscillation in this plane.

\subsubsection{Characterising frequency space}
\label{sec:char_space}

Our previous analysis has shown that the space of frequencies appears
to be well suited to identify streams. Here we will characterise this
space and how its properties depend on the form of the host
gravitational potential.

\begin{figure}
\centering
\includegraphics[width=80mm,clip]{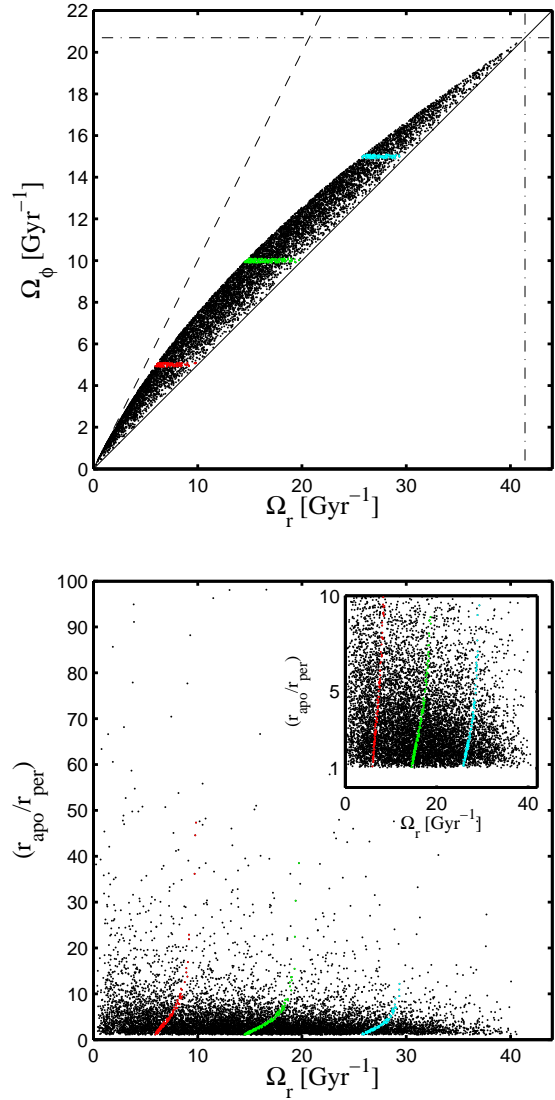}
\caption{Top: Distribution of possible orbits in a Plummer sphere in
  frequency space. The solid line corresponds to $\Omega_{r} = 2
  \Omega_{\phi}$, as in the case of orbits in a homogeneous
  sphere. The dashed line shows $\Omega_{r} = \Omega_{\phi}$, as found
  for all orbits in a point mass distribution. The left wedge of the
  distribution corresponds to circular orbits. Bottom: Relation
  between the radial frequency $\Omega_{r}$ and the orbital
  apocentre-to-pericentre ratio.}
\label{fig:plum}
\end{figure}

The angular and radial periods of an orbit are related to the respective 
frequencies through
\begin{equation}
\label{t_vs_o}
\displaystyle
T_{\phi}=2\pi/\Omega_{\phi},~~~T_{r}=2\pi/\Omega_{r}.
\end{equation}
Once a radial oscillation is completed the azimuthal angle has increased by an amount
$\Delta\phi = \Omega_{\phi}T_{r}$ or, equivalently
\begin{equation*}
\displaystyle
\Delta\phi = 2\pi\frac{\Omega_{\phi}}{\Omega_{r}},
\end{equation*}
thus
\begin{equation}
\label{os_relation}
\displaystyle
\frac{\Omega_{r}}{\Omega_{\phi}}=\frac{T_{\phi}}{T_{r}}=\frac{2\pi}{\Delta\phi}.
\end{equation}

\begin{figure}
\vspace{-0.2cm}
\centering
\includegraphics[width=80mm,clip]{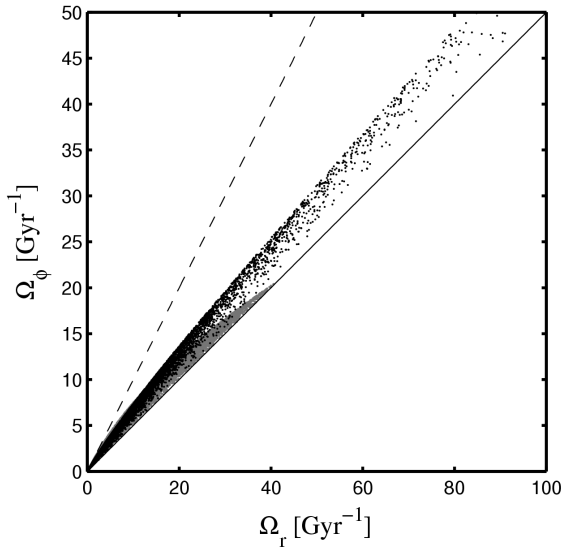}
\caption{Distribution of possible orbits in a Hernquist profile (black
  dots) in frequency space. For comparison, the distribution for a
  Plummer sphere is shown in grey. Solid and dashed lines represent as
  before the natural limits of this space, i.e.  $\Omega_{r} = 2
  \Omega_{\phi}$ and $\Omega_{r} = \Omega_{\phi}$ respectively. Note
  that because the Hernquist inner density profile is cusped orbits
  with high frequencies do not bend over to the boundary associated
  with the homogeneous sphere.}
\label{fig:hernq}
\end{figure}
In general $2\pi/\Delta\phi$ will not be a rational number and
therefore the orbit will not be closed. However there are two
gravitational potentials in which all orbits are closed, those of an
homogeneous sphere and of a point mass. Notice that any other
spherical system may be considered to have a mass distribution between
these limiting cases, i.e. its density gradient will be neither as
shallow nor as steep as in these two cases. Orbits in a
time-independent homogeneous mass distribution have $\Delta\phi=\pi$,
whereas for the Kepler problem $\Delta\phi=2\pi$. From
Eq.~(\ref{os_relation}) this implies that the frequencies of a bound
orbit in any spherical mass distribution are constrained to lie inside
the subspace defined by $\Omega_{r}/2 \leq \Omega_{\phi} \leq
\Omega_{r}$.  The extent to which this subspace is probed by orbits
depends on the exact distribution of mass.

To examine this, let us consider the Plummer potential, $\Phi$, generated by a 
density $\rho$, where
\begin{equation}
\begin{array}{ll}
\displaystyle
\label{plummer}
\Phi(r)=\frac{-GM}{\sqrt{r^{2}+b^{2}}},\\
\\
\displaystyle
\rho(r)=\frac{ 3M }{ 4{\pi}b^3}\left( 1 + \frac{ r^2 }{ b^2 }\right)^{ -5/2 }
\end{array}
\end{equation}
\citep{plum}. We sample the frequency space in this potential by
randomly drawing $10^{4}$ orbits from the distribution function
associated to the Plummer sphere with parameters $M =
10^{12}\,{\rm M_{\odot}}$ and $b \approx 22$ kpc.  The upper panel in
Fig.~\ref{fig:plum} shows the distribution of these orbits in
frequency space. The solid line and dashed lines correspond to
$\Omega_r = 2 \Omega_{\phi}$ (homogeneous sphere) and $\Omega_r =
\Omega_{\phi}$ (point mass), respectively.  Note that orbits with
small frequencies cover the whole region between these two
lines. Instead, orbits with high frequencies tend to bend over
towards the line given by the homogeneous sphere. This is because
particles on orbits with low frequencies populate the outer regions of
the system, and hence perceive a potential close to that of a point
mass. On the other hand those on high frequency orbits are generally
confined to the central regions of the system. Since the Plummer
density profile has a core these particles will feel a potential
resembling that of the homogeneous distribution. The strict cut-off at
a value of $\Omega_{r} \approx 42$~Gyr$^{-1}$ therefore corresponds to
the characteristic frequency associated to the homogeneous mass
distribution, i.e for $r << b$, a particle in the Plummer model should
have a period of $T_{r} \sim \sqrt{\frac{\pi^2 b^3}{G M}}$ which
in our case corresponds to $0.15$ Gyr, or $\Omega_{r}=42$
Gyr$^{-1}$.

For a given $\Omega_{\phi}$, (nearly) circular orbits will be those
with the largest radial period, i.e. the smallest $\Omega_r$
frequency. Thus such orbits are located on the wedge seen in the top
panel of Fig.~\ref{fig:plum}. This is also illustrated in the bottom
panel of the same figure, where we have plotted the ratio of apocentre
to pericentre distances of orbits as a function of $\Omega_{r}$. As we
can see, orbits with the lowest $\Omega_{r}$ for a given
$\Omega_{\phi}=cst$ have $r_{\rm apo}/r_{\rm per}=1$.

We now consider a system with a cuspy density distribution, namely the
Hernquist profile whose potential, $\Phi$, and density
$\rho$, are given by
\begin{equation}
\begin{array}{ll}
\label{hernq}
\displaystyle
\Phi(r)=\frac{-GM}{r+b},\\
\\
\displaystyle
\rho(r)=\frac{ bM }{ 2{\pi}r}\frac{1}{(r+b)^{3}}
\end{array}
\end{equation}
\citep{hernq}. We set $M = 10^{12}\,{\rm M_{\odot}}$ and $b \approx 22$ kpc. 
As before, we sample the available phase-space with 10$^4$ orbits. The
black points in Fig.~\ref{fig:hernq} show the resulting distribution
of orbits in frequency space.  For comparison, we overplotted the
frequencies for the Plummer profile (in grey). At the low frequencies
end both distributions overlap as expected (point mass
regime). However there is a significant difference in the region
populated by orbits with high frequencies. Due to the lack of a core
in the Hernquist density distribution, the bend over towards
$\Omega_{r} = 2 \Omega_{\phi}$, i.e. the regime of the homogeneous
mass distribution, has disappeared, and there is no upper limit to the
radial or angular frequency of motion.

\begin{figure*}
\centering
\includegraphics[width=145mm,clip]{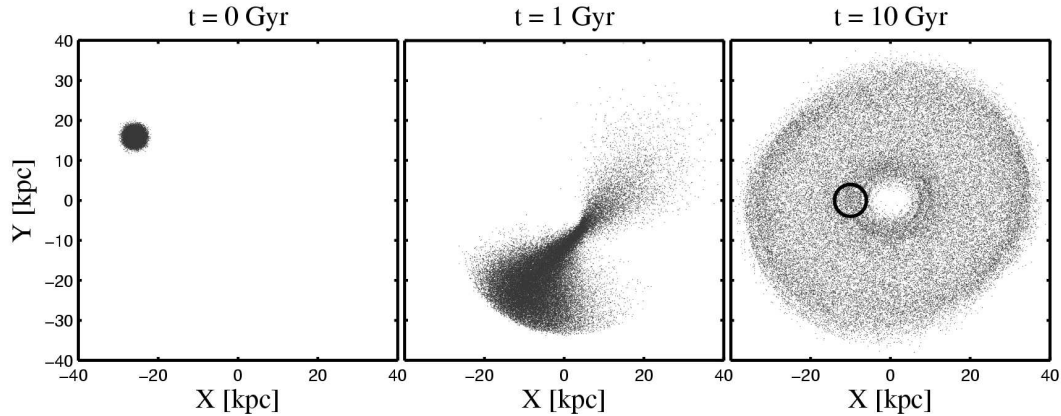}
\caption{X-Y distribution of particles from an accreted satellite at
  three different times, as indicated on each panel. The black circle
  shows the location of a ``Solar neighbourhood'' sphere.}
\label{fig:real_sp}
\end{figure*}

\begin{figure*}
\centering
\includegraphics[width=150mm,clip]{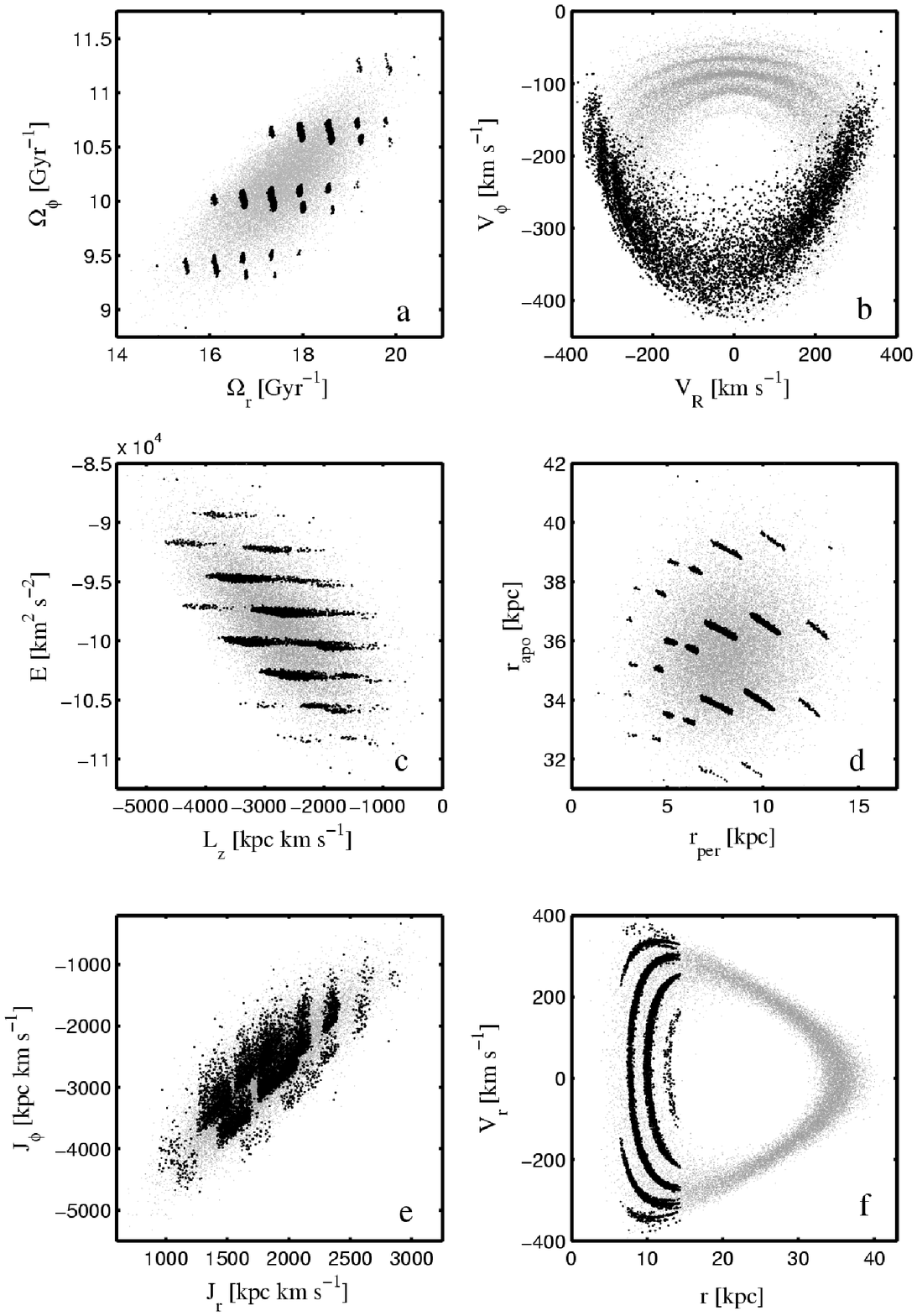}
\caption{Comparison of the various spaces commonly used to identify
  merger debris. Grey dots show the distribution of all satellite
  particles whereas the black dots represent the particles inside the
  sphere shown in Fig.~\ref{fig:real_sp}. As in the toy model example
  discussed in Section~\ref{sec:toy_model}, streams in frequency space
  (top left panel) are distributed in a regular pattern. Notice that
  the distribution of particles in frequency space is much better
  defined than in any other of the spaces considered.}
\label{fig:comp}
\end{figure*}

\subsubsection{Comparing different projections of phase-space}
\label{sec:comp}
Various spaces have been proposed to identify debris from disrupted
satellites.  For example, \citet{hz00} introduced the space of energy,
total angular momentum and its $z$-component $(E,L,L_z)$.  Later on,
\citet{h06} showed that substructure can be identified by looking at the
apocentre, pericentre and $L_{z}$ space.  In this space, streams are
represented by extended structures, roughly along a line of constant
eccentricity. In this section we will compare the distribution of
debris in these and other previously used spaces to that in
frequency space. To this end we consider the accretion of a
satellite galaxy onto a spherical host.

We represent the host with the Plummer profile discussed in the
previous section.  The satellite is assumed to be spherical and to
have an isotropic velocity ellipsoid, and is modelled with a
multivariate Gaussian in 6D with a dispersion of $\sigma_{{\bf{x}}}
\approx 1$ kpc and $\sigma_{{\bf{v}}} = 22$ km/s$^{-1}$. We set the
central particle of the satellite on an orbit with apocentre $r_{\rm
apo} \approx 35$ kpc, pericentre $r_{\rm per} \approx 8$ kpc and
inclined by 56 degrees with respect to the $z$-axis. We follow
then the evolution of the satellite for approximately $10$ Gyr.  For
simplicity, in this experiment the self-gravity of the satellite is
not considered. As a consequence, we expect to find a larger number of
streams at any given epoch, with a larger spread in orbital
parameters, in comparison to the self-gravitating case.  This is
because particles can drift apart from each other right from the
start, whereas when their self-gravity is considered, this will only
happen during successive pericentric passages where they are
(typically) stripped from the parent satellite.

In Fig.~\ref{fig:real_sp} we show the spatial distribution of
satellite particles at three different times. To study the
distribution of particles in various projections of phase-space we
focus on those inside an sphere of radius $4$ kpc centred at 
$r \approx 10$
kpc at the final time, as shown in the rightmost panel of this Figure.

The results are shown in Fig.~\ref{fig:comp} for six
different spaces. Grey points represent all the particles of the
satellite whereas black points only those inside the sphere of
interest. In the top left corner (panel a) we show their projection in
frequency space. As in the toy model discussed in Section
\ref{sec:toy_model}, for a given $\Omega_{\phi} = cst$ we find
multiple streams in $\Omega_{r}$. Note that there are typically more
streams at constant $\Omega_{\phi}$ than at constant $\Omega_{r}$.
This is because $T_{r} \leq T_{\phi} \leq 2T_{r}$, and hence mixing
occurs faster in the $r$-direction.

Notice that some of the groups of particles found at given
$\Omega_{r}$ and $\Omega_{\phi} \approx cst$ may be decomposed into
smaller structures.  This can be understood from the radial velocity
vs. radial distance plot shown in the bottom right panel of
Fig.~\ref{fig:comp}. Streams with pericentres inside the sphere under
consideration, appear in frequency space as a single compact
structure. For the rest we observe two separate structures, each of
these associated to particles located just before or after their
corresponding pericentre.

The top right panel of Fig.~\ref{fig:comp} shows the distribution of 
particles in a projection of velocity space. Although several features 
are visible, the individual streams are clearly less well separated than 
in panel a). This is also the case for the radial velocity vs. radial 
distance plot (panel f). 

The two middle panels of Fig.~\ref{fig:comp} show the distribution of
particles in the $E-L_{z}$ (left) and apocentre vs. pericentre (right)
spaces.  The number of structures in the $E-L_{z}$ space is smaller
and slightly less sharp than in frequency space. On the other hand,
streams are well detached from each other in the apocentre
vs. pericentre space. Finally, in the bottom left panel we show the
distribution of particles in the space of actions, $J_{r}$ vs.
$J_{\phi}$. As in the $E-L_{z}$ space, distinction of the various
streams is less clear in this space, and hence it appears to be less
suitable for finding individual debris streams. However, it is important
to note that we are only considering two of the three available actions.
By projecting the particle's distribution into the principal directions
of action space it is also possible to find a well defined
 distribution of streams (see Figure 7 of MB08).

This analysis shows that frequency space is a very suitable space to 
identify streams associated to accretion events. Moreover, as we
explained in  Section \ref{sec:toy_model}, the way streams are distributed 
in this space can allow us to estimate quantities such as time since 
disruption. This last characteristic makes the space of orbital 
frequencies more appealing than the other spaces previously considered.

\subsubsection{Estimating the time of accretion}
\label{sec:time_acc}

As discussed before the separation between adjacent streams along each
direction of the space of orbital frequencies yields a direct estimate
of the time since disruption, e.g. a satellite accreted 10 Gyr ago should
give rise to streams separated by a scale $\delta{\Omega_{r}} =
\delta{\Omega_{\phi}} = 2\pi/10$ Gyr$^{-1}$. 

By considering information from the distribution of particles not only 
in frequency but also in angle space MB08 developed a powerful method that, 
by a process of iteration over a guess potential, allows to estimate time 
of disruption and to pin down the true underlying potential.
 
In this section we explore a different method which, only based on the 
regularity seen in frequency space, provides an accurate estimate
of time since disruption. As an example we focus on the streams shown  
in panel a) of Fig.~\ref{fig:comp}.  Because there is a unique scale 
(all streams are separated by the same amount) 
a straightforward way to compute the separation between adjacent streams
 is to use Fourier analysis techniques. We proceed by first creating 
an image from the scatter plot in frequency space, to which we then apply 
a Fourier transform. In the final step we compute the power spectrum. 
To obtain an image from our data we grid the frequency space with a 
regular $N \times N$ mesh of bin size $\Delta$ in both $\Omega_{r}$ and
$\Omega_{\phi}$ directions.  In this way, the number of particles in
the $(n_{r},n_{\phi})$ bin of the grid is $h(n_{r},n_{\phi})$. We
apply a two dimensional discrete Fourier transform to this image:
\begin{multline}
\label{Fourier_transf}
\displaystyle
H(k_{r},k_{\phi})=\sum^{N-1}_{n_{r}=0}
\sum^{N-1}_{n_{\phi}=0}\exp(2\pi i k_{r} n_{r} / N)
\\ \exp(2\pi i k_{\phi} n_{\phi} / N) h(n_{r},n_{\phi}),
\end{multline}
with $k_{r},k_\phi=-\frac{N}{2},\dots,\frac{N}{2}-1$.

\begin{figure}
\centering
\includegraphics[width=64mm,clip]{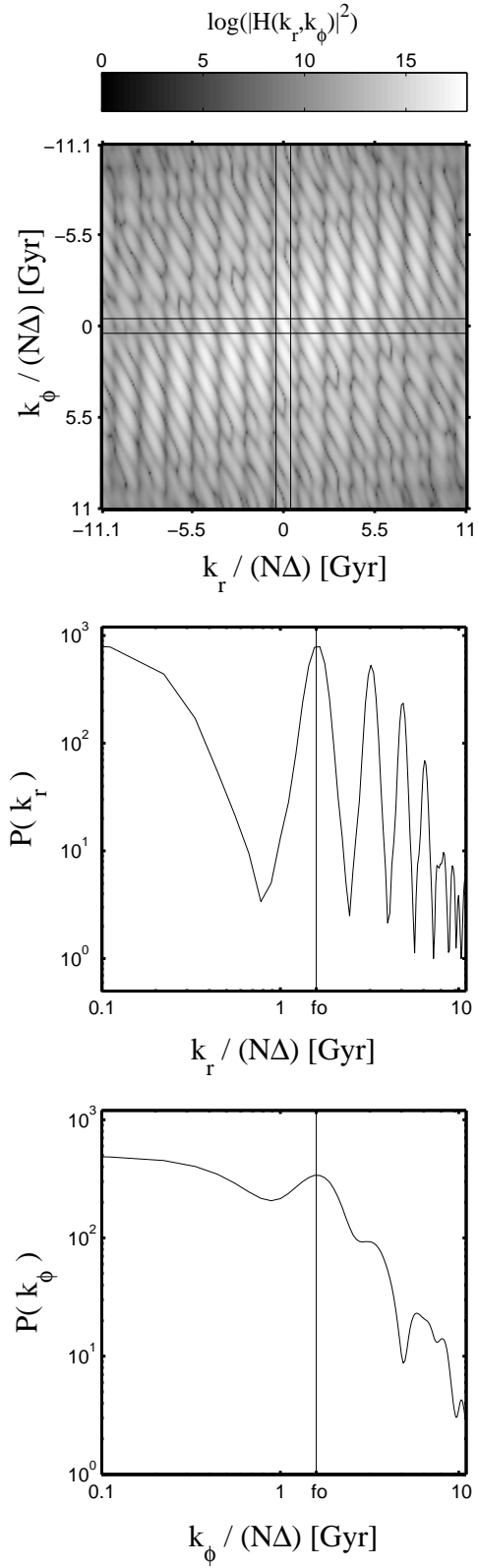}
\caption{Top panel: Fourier transform of the distribution of streams
  shown in the top left panel of Fig.~\ref{fig:comp}. The black lines
  denote the regions selected in the Fourier image to compute the 1-d
  power spectra, i.e. around $k_{\phi}=0$ (middle panel) and $k_{r}=0$
  (lower panel). The black vertical lines in these panels show the
  location of the peak with the highest power in the spectrum.}
\label{fig:ps_hr}
\end{figure}

In the top panel of Fig.~\ref{fig:ps_hr} we show the image of the 
panel a) of Figure ~\ref{fig:comp} in Fourier space. This has been
computed on a grid of 200 by 200 elements.  The axes represent the
wavenumbers in each direction ($k_r/(N\Delta)$, $k_{\phi}/(N\Delta)$) while 
the colour coding show the square value of each of the Fourier components
$|H(k_r,k_{\phi})|^2$.  As we can see from this image, and as expected
from our previous discussion, a regular pattern is present. To measure
the characteristic scale we consider two different ``slits'' (of 4 bins
width) of the Fourier image around $k_{r}=0$ and
$k_{\phi}=0$. These regions are delimited by the black lines in the
top panel of Fig.~\ref{fig:ps_hr}. We can now compute the
one-dimensional power spectrum of the image along each direction as
\begin{equation}
\begin{array}{llll}
\displaystyle
\label{ps_estimation}
P(0)=\frac{1}{N^2} \left|H(0,0)\right|^2, \\
\\
\displaystyle
P(k_{r})=\frac{1}{N^2}\left[\left|H(k_{r},0)\right|^2+\left|H(-k_{r},0)\right|^2\right]\\
\\
$for $ k_{r}=1,\dots,\left(\frac{N}{2}-1\right),\\
\\
\displaystyle
P(N/2)=\frac{1}{N^2} \left|H(-N/2,0)\right|^2 \\
\end{array}
\end{equation}
and analogously for $P(k_{\phi})$. Recall that the one-dimensional power spectrum we
compute is, in practice, an average over the slits considered along 
the $k_r$ and $k_\phi$ directions. 

The results are shown in the middle and the bottom panels of
Figure~\ref{fig:ps_hr}, for $P(k_{r})$ and $P(k_{\phi})$,
respectively. The scales present in the power spectrum $P(k_{r})$ are
associated to separations in the $\Omega_{r}$ direction (since they
correspond to $k_{\phi}=0$). The large power for wavenumbers close to
zero, i.e. $P(0)$ represents the average power over the whole
image. All other peaks in the spectrum are related to waves resonating
with our image. In particular the first peak, with the highest power,
is associated to a wavenumber of $f_{0}=1.59$ Gyr, and is denoted by
the vertical black line in the spectrum. The corresponding wavelength
of this peak is $0.62$ Gyr$^{-1}$ $\approx 2\pi/10$ Gyr$^{-1}$. This
is in fact the separation we would have estimated using
Eq. (\ref{sep}) and corresponds to an estimated time since accretion
of $\tilde{t}_{\rm acc}=2\pi f_{0} \approx 10$ Gyr . The rest of the
peaks in the spectrum are associated to the harmonics of this
wavenumber, i.e., $2f_{0}$, $3f_{0}$, etc.

The situation is similar when we consider a cut around $k_{r}=0$. In
this case the scales present in the power spectrum are associated to
separations in the $\Omega_{\phi}$ direction. We find the peak
with the highest amplitude to be at the same wavenumber $f_{0}$ and it is
possible to distinguish some of its harmonics. However, due to the
smaller number of streams we have in the $\Omega_{\phi}$ direction,
the peaks are less clearly defined than for $P(k_r)$.

The Fourier method developed in this section may be considered to be
very complementary to that introduced by MB08. Even if the
gravitational potential is only approximately known, streams in
frequency space will still distributed on a lattice (see below and
Sec. 4.3 of MB08). This implies that a Fourier analysis technique will
(always) provide an estimate of the time since accretion. Instead, in
the presence of a very clear set of streams MB08's method can be used
not only to provide a more accurate estimate of the time since
accretion but also to narrow down the true underlying potential.

\section{Frequency space for time dependent potentials}

In a hierarchical universe, the gravitational potentials associated to
galaxies are expected to have changed over time.  Under these
circumstances, quantities such as the energy, apocentre and pericentre
and also the frequencies of individual orbits will no longer be
conserved.  It is therefore of special interest to study how a time
dependent potential affects the structure of debris in frequency
space.

To this end, we consider a simple model of a satellite accreted onto a
Plummer sphere whose mass varies according to
\begin{equation}
\label{m_t}
M(t)=M_{0}\exp(t/t_{\rm scale}) 
\end{equation}
where we explore several values for the time scale, $t_{\rm scale}$.
The growth in mass will lead to a shrinking of the possible orbits,
and be manifested in an increase in the frequencies and acceleration
experienced by the particles with time.

\subsection{Time evolution of the frequencies and the rate of change of the potential}
\label{sec:time_evol_frq}

To derive the evolution of the frequencies of motion for an orbit in a
time-dependent mass distribution such as that given by
Eq.~(\ref{m_t}), we proceed as follows. At each step of this orbit's
time-integration, we take the instantaneous position and velocity as
initial conditions to compute an orbit in the instantaneous (frozen)
associated potential. Then, for this orbit we derive an instantaneous
set of frequencies.

We will consider the evolution of two different orbits: an initially
outer orbit with an apocentre of $r_{\rm apo} \approx 64$ kpc 
and pericentre of
$r_{\rm per} \approx 19$ kpc, and a second inner orbit with 
$r_{\rm apo} \approx 25$
 kpc and $r_{\rm per} \approx 5$ kpc. The initial values of 
the frequencies 
of the outer orbit are $\Omega_{r} \approx 8$ Gyr$^{-1}$ and 
$\Omega_{\phi} \approx 6$ Gyr$^{-1}$, whereas for the inner 
one they are $\Omega_{r} \approx 25$ Gyr$^{-1}$ and 
$\Omega_{\phi} \approx 13$ Gyr$^{-1}$. Both orbits 
are launched from their apocentre.

\begin{figure}
\centering
\includegraphics[width=84mm,clip]{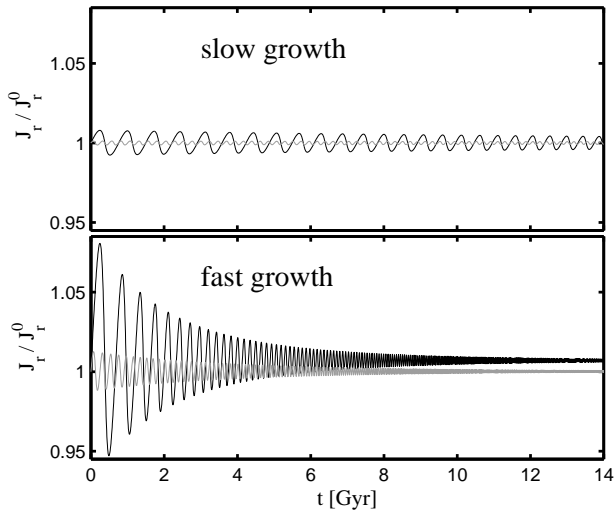}
\caption{Time evolution of the radial action of two different orbits
  in a very slow (top panel) and a very fast (bottom panel) time
  varying Plummer potentials. The black and light-grey curves
  correspond to an outer and an inner orbit respectively. The observed
  oscillatory behaviour is related to the orbital motion.}
\label{fig:Jr_evol}
\end{figure}

Let us first consider a slowly varying potential, where $t_{\rm scale}
= 30$ Gyr. We follow the evolution of the orbit over $14$ Gyr. The top
panel in Fig.~\ref{fig:Jr_evol} shows the time evolution of the radial
action $J_{r}$ for this case, where black and light-grey lines in this
Figure correspond to the outer and the inner orbit respectively. This
Figure shows that the instantaneous action $J_{r}$ displays an
oscillatory behaviour\footnote{Note that due to the spherical nature
of the potential the other two actions, associated with the angular
momentum, remain constant in time.}, with a period that corresponds to
that of the orbit.

This is not surprising since under adiabatic variations of the
potential it is the time average of the actions that remains constant
\citep[see][]{goldstein,ws}. In general, for a given orbit the
amplitude of the oscillation depends on the time scale on which the
potential grows. By increasing $t_{\rm scale}$ the evolution is more
adiabatic and the amplitude of the oscillation is decreased.

In order for an orbit to be in a condition of adiabatic invariance it
is necessary that $t_{\rm orb} \ll t_{\rm scale}$, where $t_{\rm orb}$ is the
orbital period. Since inner orbits have shorter orbital periods, we
expect those to behave ``more adiabatically" compared to outer orbits.
This is demonstrated in the top panel of Fig.~\ref{fig:Jr_evol}, where
the amplitude of the oscillation of $J_{r}$ is smaller for the inner
orbit.

For the outer orbit in the rapidly growing potential, $t_{\rm
scale} = 3$ Gyr, the amplitude of oscillation is larger, especially at
early times when the orbit is farther away from the adiabatic regime
(since its initial period, $T_{r}=0.78$, is comparable to $t_{\rm
scale}$). As the mass of the system grows in time, which leads to the
deepening of the potential well, orbits are continuously driven
towards deeper regions of the well, implying that the corresponding
orbital periods decrease with time. Therefore what we observe as the
damping of the amplitude of oscillation is a transition from a
non-adiabatic towards an adiabatic regime.

\begin{figure}
\centering
\includegraphics[width=84mm,clip]{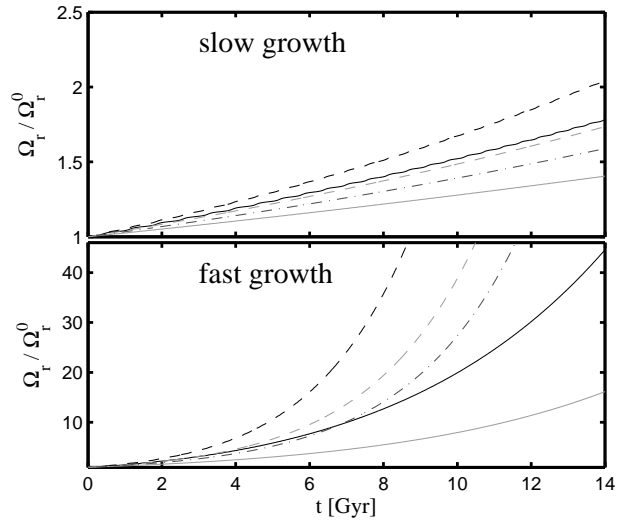}
\caption{Time evolution of the radial frequency for two different
  orbits in a very slow (top panel) and a very fast (bottom panel)
  time varying Plummer potentials. The black and light-grey curves
  correspond to an outer and an inner orbit respectively. The dashed
  lines represent the time evolution of the normalised energies of
  each orbit $E/E_{0}$, while the dotted-dashed line shows the time
  evolution of the mass of the host $M/M_{0}$.}
\label{fig:Or_evol}
\end{figure} 

In Figure~\ref{fig:Or_evol} we focus on the evolution of the radial
frequency $\Omega_{r}$ (the angular frequency $\Omega_{\phi}$ depicts
a qualitatively similar behaviour). The top and bottom panels of this
Figure correspond, respectively, to the slowly and to the rapidly
varying potentials.  Note that in contrast to the actions, the
frequencies do change with time, and that their evolution appears to
follow closely that of the potential. For comparison, we plot in
Fig.~\ref{fig:Or_evol} the time evolution of the normalised energy of
each orbit $E/E_{0}$ (dashed lines), and of the mass of the system
$M/M_{0}$ (dotted-dashed line).

The exact relation between the evolution of the frequencies and that
of the potential can not be derived analytically in the general
case. However, some insights may be obtained from two simple cases:
the homogeneous sphere and the Kepler potentials.

For the homogeneous sphere the radial period $T_{r} \propto
\rho^{-\frac{1}{2}}$ or $T_{r} \propto M^{-\frac{1}{2}}$ \citep{bt},
therefore $\Omega_{r} \propto M{^\frac{1}{2}}$. Hence, for
$M(t)\propto \exp(t/t_{\rm scale})$ we obtain
\begin{equation}
\Omega_{r}(t) \propto \exp(t/2t_{\rm scale}).
\end{equation} 
Therefore, the frequencies evolve in time with the same functional
form as the potential, but on a time scale that is twice as
long. Consequently, their evolution is slower.

\begin{figure}
\vspace{0.2cm}
\centering
\includegraphics[width=80mm,clip]{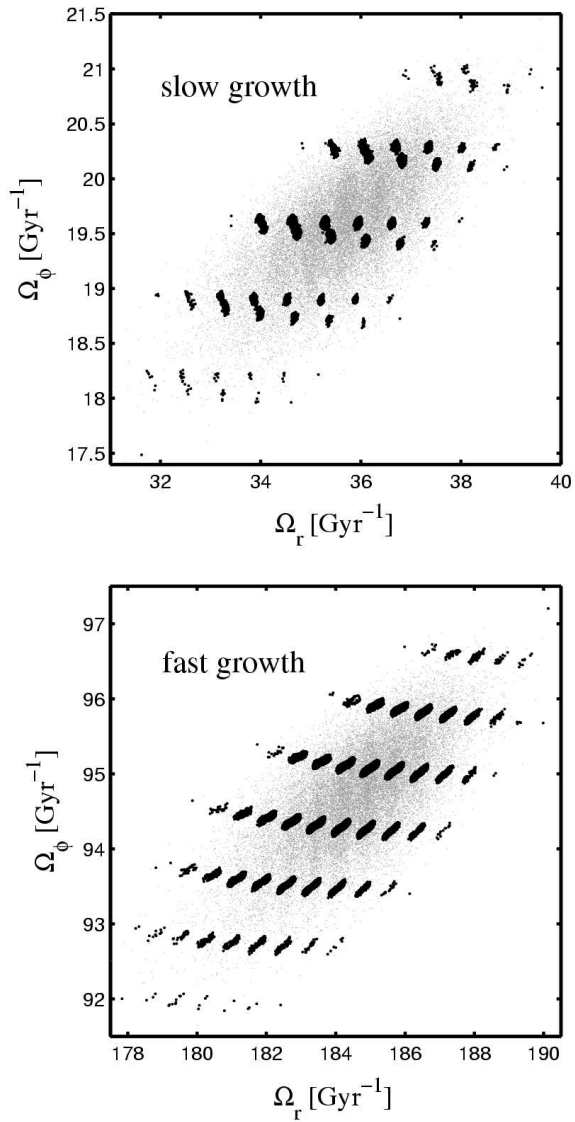}
\caption{Distribution of particles in frequency space located inside a
  sphere of 4 kpc radius at $\approx 10$ kpc from the centre of the
  time-dependent Plummer sphere after 10 Gyr of evolution. The
  timescales for the mass growth of the Plummer sphere are
  $t_{\rm scale}= 12$ Gyr (top panel) and $t_{\rm scale}= 3$ Gyr 
  (bottom panel). Note that structure in frequency
  space is not destroyed by the time varying nature of the potential.}
\label{fig:plum_tevol}
\end{figure}

In a Kepler potential the radial orbital frequency of a particle can 
be expressed in terms of its actions as \citep[][Appendix E]{bt}
\begin{equation}
\label{o}
\displaystyle
\Omega_r=\frac{\left(GM\right)^{2}}{\left(J_{r}+L\right)^{3}}
\end{equation}
and therefore
\begin{equation}
\label{osol}
\displaystyle
\Omega_r(t) \propto \exp(2t/t_{\rm scale}).
\end{equation}

Again, in this example we find that frequencies are evolving in a
similar fashion as the potential. Therefore a very external orbit in
the Plummer sphere should initially evolve at a rate similar to that
given by Eq.~(\ref{osol}) (as for the Kepler potential). On the other
hand, inner orbits should evolve at a rate similar to that of the
homogeneous sphere. Thus, we can already see that orbits in a given
potential will evolve at different rates, depending on their initial
location in phase-space.

\subsection{Structure in frequency space}
\label{sec:streams_t_evol}

In this section we address how the debris of a satellite is
distributed in frequency space in the time-dependent Plummer potential
discussed above.

We ran simulations with the same satellite and orbital initial
conditions as described in Section \ref{sec:comp}. We followed its
evolution in the Plummer potential with two different timescales,
namely $t_{\rm scale}=3$ Gyr and 12 Gyr. As before, we
do not include the self-gravity of the satellite.

Fig.~\ref{fig:plum_tevol} shows the distribution of particles inside a
sphere of $4$ kpc radius located at $ \approx 10$ kpc from the centre of the
system after 10 Gyr of evolution. The top panel presents the results
for $t_{\rm scale}=12$ Gyr while the bottom panel corresponds to
$t_{\rm scale}=3$ Gyr. As we can see from this Figure, even in a time
dependent system, satellite debris at a given spatial location is
distributed in a regular pattern of patches in the space of orbital
frequencies.

In the simulation where the time evolution is slow (i.e.,
$t_{\rm scale}=12$ Gyr) some of the structures may be decomposed into two
clumps. As we explained in Section~\ref{sec:comp}, such structures
appear when neither the apocentre nor the pericentre of the orbits
fall inside the sphere under consideration. In contrast, when the
potential evolves quickly the particles in each stream are all
distributed in compact single structures. This is because in this case
the orbits have shrunk so much after $10$ Gyr that all streams present
in this volume have their apocentres inside the chosen sphere.

\begin{figure}
\centering
\includegraphics[width=80mm,clip]{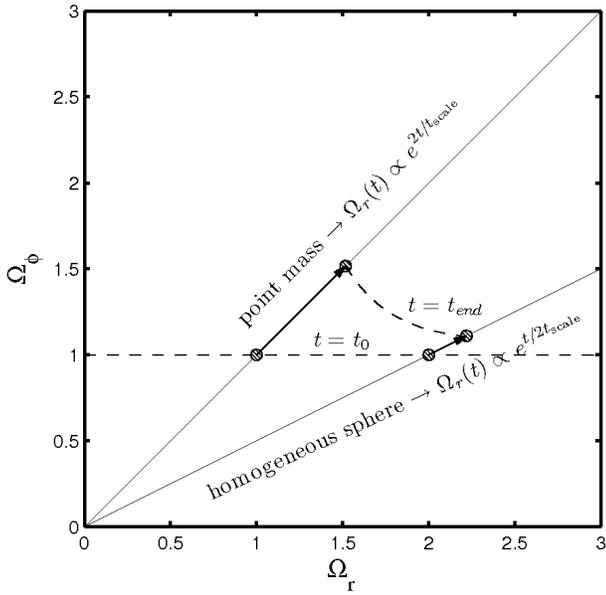}
  \caption{Schematic of the time evolution of the frequencies of two
particles and their dependence on location in frequency space.  At
initial time, $t=t_{0}$, both particles have the same initial
$\Omega_{\phi}$ but different $\Omega_{r}$. The frequency of the
particle closer to the homogeneous sphere regime evolves at a slower
rate than that which is closer to the point mass distribution. This
results in a lattice that has curvature in frequency space for those
particles which satisfy $\int_{t_{0}}^{t_{end}} \Delta\Omega_{r}(t) dt
= 2\pi N$, similar to that seen in Figure~\ref{fig:plum_tevol}.}
\label{fig:esquema}
\end{figure}

Comparison to the case in which the potential is fixed
(Fig.~\ref{fig:comp}) shows that streams in the $\Omega_{r}$ direction
are not exactly distributed along a line of $\Omega_{\phi}=cst$ but
rather on a line with a given curvature. This is due to the fact that
orbits with different initial frequencies evolve at different rates
(as discussed in the previous section).  The amount of curvature
observed in this space (e.g. Figure \ref{fig:plum_tevol}) depends on
the rate of evolution of the potential. This is schematically depicted
in Figure~\ref{fig:esquema}.  Particles located close to the
homogeneous sphere regime, will evolve at rates $\sim (2 t_{\rm
scale})^{-1}$, while those with orbits closer to the point mass
distribution regime will evolve at rates more similar to $\sim
2/t_{\rm scale}$, i.e., four times faster.  Note that since the
particles in the Plummer sphere occupy a smaller region in frequency
space than that delimited by these two regimes, the difference in
their rate of evolution will typically be smaller than the estimate
just provided. If by the final time $t_{end}$, the particles
satisfy the condition $\int_{t_{0}}^{t_{end}} \Delta\Omega_{r}(t) dt =
2\pi N$ their associated streams will be located along a curve roughly
as depicted in this Figure.

Note that the extent of the distribution of particles in frequency
space has also changed, but since this is also dependent on the
initial internal properties of the satellite, it cannot be used to
directly measure the evolution of the potential.

We may conclude from these experiments that the time evolution
modelled here for the host potential does not destroy the coherence of
satellite debris in frequency space. Furthermore, since the
frequencies evolve at different rates even for particles with same
origin, it may be possible to recover the evolution of the host
potential by measuring the curvature of the lines over which the
streams are distributed in frequency space.  Note that this implies
that the host's evolution has left its imprints in the present day
orbits of accreted stars. This is in contrast to the results by
\citet{p06} \citep[see also][]{wrck}, who find that present-day
observables only constrain the present day mass distribution of the
host, independently of its past evolution.

\begin{figure}
\centering
\includegraphics[width=64mm,clip]{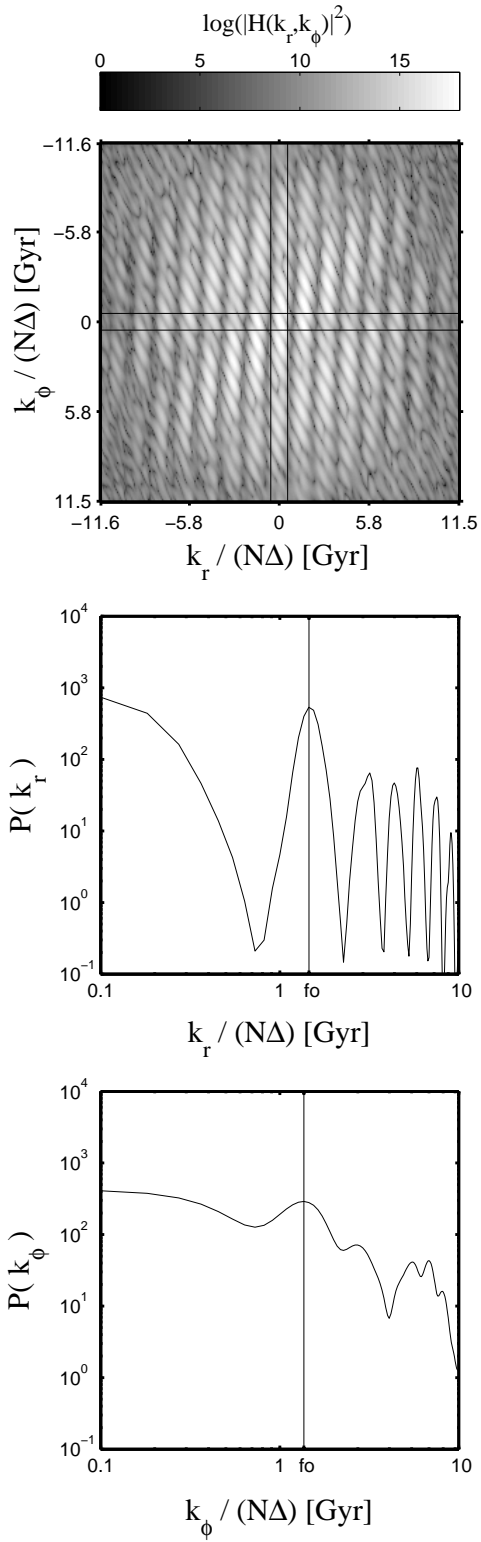}
\caption{Top panel: Fourier transform of the distribution of streams
  shown in the top panel of Fig.~\ref{fig:plum_tevol}. As in
  Fig.~\ref{fig:ps_hr} the middle and central panels depict the 1-d
  power spectra along the $k_r$ and $k_\phi$ directions,
  respectively. The location of the first peak (denoted by the
  vertical lines in these panels) can be used to estimate the
  accretion time of the satellite.}
\label{fig:ps_t40}
\end{figure} 

\subsection{Estimating the time of accretion}
\label{sec:time_acc_te}

The top panel in Fig.~\ref{fig:ps_t40} shows the results of applying a
two dimensional Fourier transform to an image created from the
distribution of streams shown in the top panel of
Fig.~\ref{fig:plum_tevol}. This corresponds to the satellite accreted
$10$ Gyr ago onto a host whose mass increases on $t_{\rm scale}=12$
Gyr. As before we compute the power spectrum along the $k_{\phi}$ and
$k_{r}$ directions. The middle panel shows $P(k_{r})$, and the black
line here denotes the location of the peak with the highest amplitude,
having a wavenumber of $f_{0} = 1.46$ Gyr. From this spectrum we would
estimate the satellite was accreted $9.2$ Gyr ago. When $P(k_\phi)$ is
considered, the highest amplitude peak is located at a wavenumber
$f_{0}=1.36$ Gyr, corresponding to an estimated time of accretion of
$8.6$ Gyr. Therefore, in this example the analysis of the power
spectrum suggests different times of accretion depending on whether
$\Omega_{r}$ or $\Omega_{\phi}$ are considered. Both values are
reasonably close to the actual accretion time ($10$ Gyr ago), but the
direction associated to $\Omega_{r}$ appears to provide a better
estimate.

\begin{figure}
\centering
\includegraphics[width=64mm,clip]{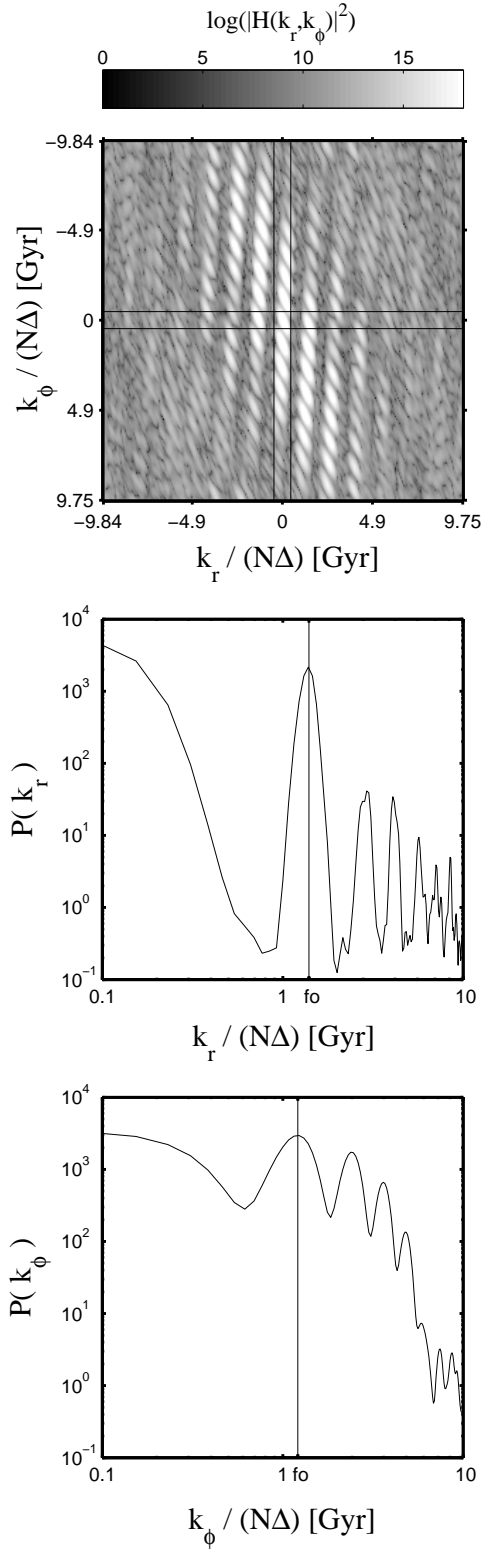}
  \caption{As in Fig.~\ref{fig:ps_t40}, now for the distribution of
  streams shown in the bottom panel of Fig.~\ref{fig:plum_tevol}.}
\label{fig:ps_t10}
\end{figure}

Fig.~\ref{fig:ps_t10} shows the same result but now for the satellite
accreted onto a host potential that evolves on $t_{\rm scale}=3$
Gyr. $P(k_{r})$ peaks at a wavenumber $f_{0}=1.4$ Gyr, as shown by the
black line in the middle panel of this Figure while $P(k_\phi)$ peaks
at a wavenumber $f_{0}=1.21$ Gyr. The corresponding estimated times of
accretion $\tilde{t}_{\rm acc}$ are $8.8$ and $7.6$ Gyr ago,
respectively. Note that even though the potential evolves on a very
short timescale, the estimated accretion times only differ by $\sim
15-25$\% from the true value.

We may estimate the relation between the true time since accretion
$t_{\rm acc}$ and $\tilde{t}_{\rm acc}$ for a general time dependent
potential as follows. Let us consider two particles that at final time
are found in adjacent streams. As discussed before, the condition for
this to happen is $\Delta\theta_{i}(t_{\rm acc}) =\int_{0}^{t_{\rm
acc}} \Delta\Omega_{i}(t) dt = 2\pi$, for $i = r,~\phi$. This
implies
\begin{equation}
\label{eqn:int} 
\displaystyle \int_{0}^{t_{\rm acc}}
\Delta\Omega_{i}(t)~dt = \Delta \Omega_{i}(t_{\rm acc})~\tilde{t}_{\rm acc}.
\end{equation} 
We may use the second mean value theorem for integration
\citep{courant}, i.e.  $\int_{a}^{b}
G(t) \varphi(t) dt = G(a) \int_{a}^{x} \varphi(t) dt + G(b) \int_{x}^{b}
\varphi(t) dt$. If we take $\varphi(t) = 1$ and $G(t) = \Delta
\Omega_{i}(t)$, then
\[\int_{0}^{t_{\rm acc}} \Delta\Omega_{i}(t)~dt = \Delta\Omega_{i}(0)t_{\rm g} 
+ \Delta\Omega_{i}(t_{\rm acc}) (t_{\rm acc} - t_{\rm g}),\]
where $t_{\rm g} \in (0,t_{\rm acc})$. Therefore in Eq.~(\ref{eqn:int})
\begin{equation}
\tilde{t}_{\rm acc} = t_{\rm acc} - t_{\rm g} 
(1 - \Delta\Omega_{i}(0)/\Delta\Omega_{i}(t_{\rm acc})),
\end{equation}
which shows that our estimate is always a lower limit to the actual
time since accretion. Its accuracy depends on the particular form of
the gravitational potential and on its evolution (through
the quantities $\Delta\Omega_{i}(t_{\rm acc})$ and $t_{\rm g}$).

We have also performed an experiment in which an axisymmetric thin
disk was grown adiabatically inside an initially spherical (Plummer)
mass distribution. Therefore, in this case, both the shape and the
depth of the gravitational potential have changed with time. We find
that also under such conditions, we are able to estimate the time of
accretion with similar accuracy.

\section{Full N-body case}
\label{sec:nbody}

The previous analysis has shown that frequency space is particularly
well suited for identifying streams from mergers, at least for the
idealised potentials considered thus far. We now explore a more
realistic case, namely that of a minor merger with a live host. We use
one of the simulations of \citet{vh08} (hereafter VH08), that model
the formation of a thick disk via the merger of a relatively massive
satellite onto a pre-existing thin disk. Two important physical
processes take place during the merger which may affect the structure
of the debris in frequency space. First, the satellite suffers
significant dynamical friction. Secondly, the host system responds
strongly to the perturber, resulting in a disk that has been
significantly tilted and heated. This implies that, even though the
total mass of the system is conserved, its distribution evolves in
time. The VH08 simulation that we consider here corresponds to a two
component satellite (stars + dark matter) launched on a $30\,^{\circ}$
inclination orbit with respect to the disk from a distance of
approximately $84$ kpc ($\sim 50$ disk scale lengths). The host
consists of a (live) dark halo and a (live) thin disk. The total (and the stellar) 
mass ratio between the satellite and the host is $20\%$. After $4.5$
Gyr of evolution, the satellite is fully disrupted and has deposited
debris in the disk of the host. The host disk is significantly
thickened by this process and shows characteristics typical of thick
disks.

\subsection{Computing the frequencies}
\label{sec:comp_freq}

The remnant system modelled by VH08 does not have a separable
gravitational potential which prevent us from deriving explicitly the
action-angles variables using
Eq.~(\ref{hamilton_prim})-(\ref{time_evol}). Nevertheless several
methods have been developed to obtain the frequencies of orbits in
general potentials. Here we focus on the spectral analysis approach,
introduced by \citet{bs82}, which relies on the numerical integration
of the equations of motion in a given potential. The basic idea is to
perform a Fourier transform of the time series ${\bf{x}}(t)$ (i.e.,
the orbit), and then to derive the frequencies of motion.

To obtain a reliable Fourier spectrum the orbits must be integrated
for a very large number of periods. However, in the problem under
consideration the orbits are not stationary, i.e. the frequencies are
time dependent and so this approach cannot be applied directly using
the orbits from the N-body simulation. This is why we have used the
final output of the simulation to define a set of initial conditions
which we integrate in a fixed (static) potential. This potential should 
ressemble that of our simulation, and for simplicity we have taken a two 
component model, consisting in a Miyamoto-Nagai disk \citep{mi-na}
\begin{equation}
\Phi_{\rm disk}=-\frac{GM_{\rm disk}}{\sqrt{R^{2}+(a+\sqrt{z^{2}+b^{2}})^{2}}},
\end{equation}  
and NFW dark matter halo \citep{nfw} 
\begin{equation}
\Phi_{\rm halo}=-\Phi_{0}\frac{r_{s}}{r}\log\left(1+\frac{r}{r_{s}}\right).
\end{equation}
We have set the parameters of the model to $M_{\rm disk}=2.5 \times
10^{10}\,{\rm M_{\odot}}$, $a=1.8$ kpc and $b=0.6$ kpc; $\Phi_{0}=1.05 \times
10^5$ km$^2$~s$^{-2}$ and $r_{s}=14.1$ kpc. These choices lead to a potential
energy distribution that differs by at most 5 per cent from the true
potential up to a radius of $50$ kpc on the $z=0$ plane.
This radius is large enough to enclose the apocentres of
99\% of the particles presently found in a volume resembling the Solar
neighbourhood. We thus integrate the orbits of the stellar particles located in
such a volume for approximately $100$ orbital periods and compute
their frequencies using the code developed by \citet{daniel}.

Note that the results are not strongly dependent on our particular
choice of the potential, since substructure in integrals of motion
space \citep{hz00,h06} and in frequency space (MB08) are both robust
to small differences in the mass distribution.  This is because we
always focus on small volumes in space and hence small changes in the
potential essentially act as a zero point offset, affecting all
particles present in this volume in the same way.

\subsection{Results}

\begin{figure*}
\centering
\includegraphics[width=120mm,clip]{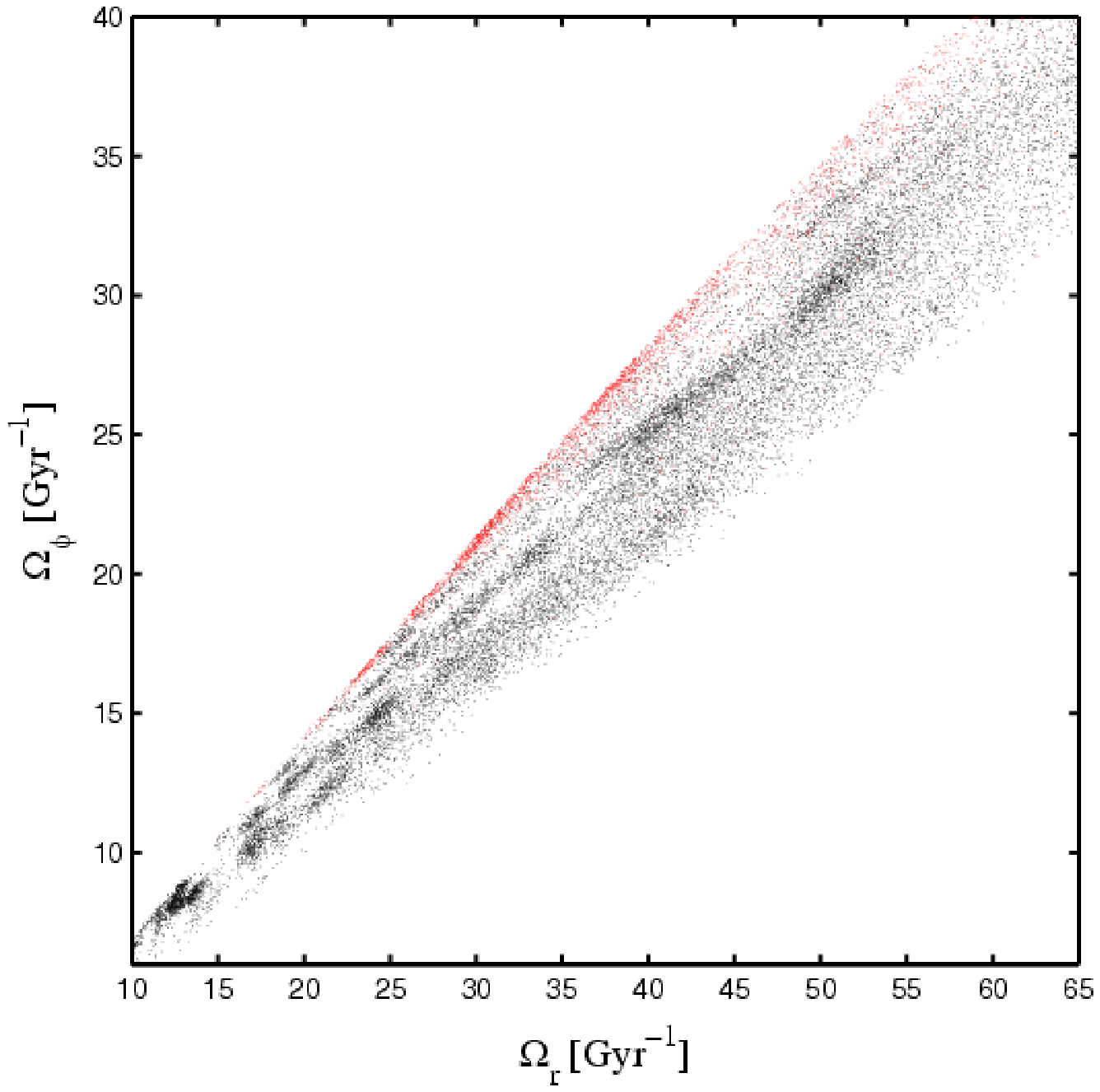}
\includegraphics[width=160mm,clip]{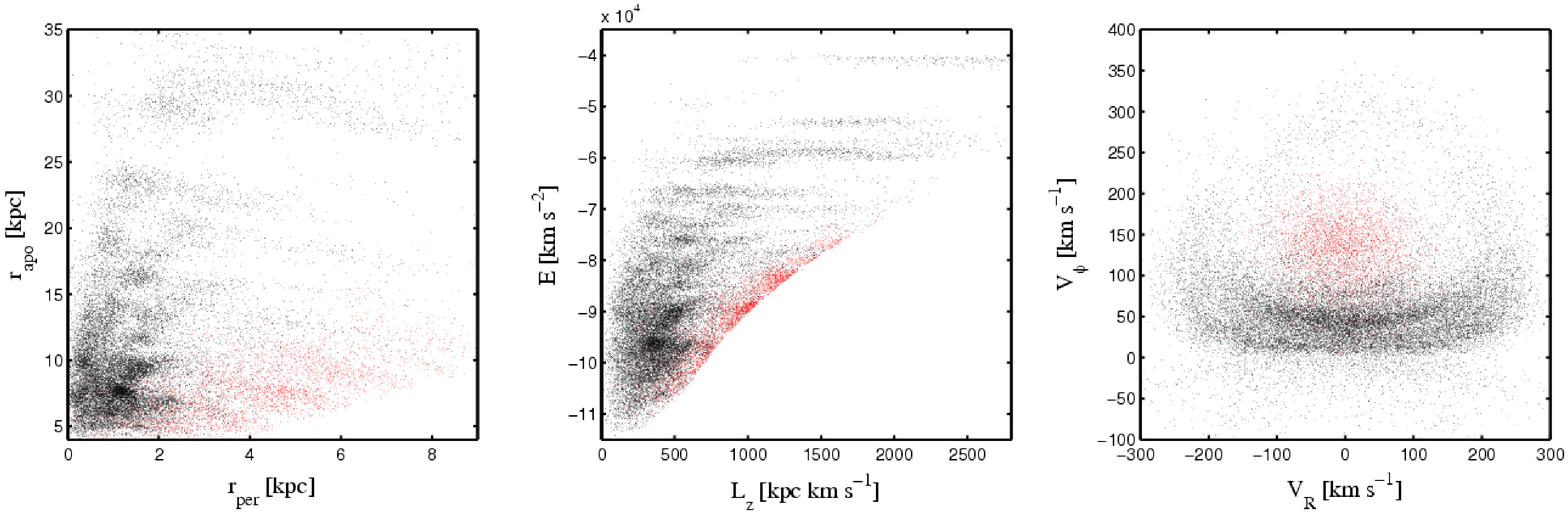}
\caption{Distribution of stellar particles inside a sphere of 4 kpc
  radius located at 8 kpc from the centre of the remnant of a 5:1
  merger between a satellite and a disk galaxy, after 4.5 Gyr of
  evolution. In all panels, the black dots represent particles from
  the satellite and red dots particles from the disk. The top panel
  shows the distribution of particles in frequency space, whereas the
  bottom panels correspond to the apocentre vs. pericentre space
  (left), the $E-L_{z}$ (middle) and the $V_{R}-V_{\phi}$ (right)
  projections. Notice that streams are much better defined in
  frequency space than in these other projections.}
\label{fig:streams}
\end{figure*}

In Fig.~\ref{fig:streams} we present the distribution of stellar
particles inside a sphere of $2$ kpc radius located at $8$ kpc from
the centre of the remnant galaxy in some commonly used spaces to
identify substructure. Here the red particles represent those
originally present in the disk, while those from the satellite are
colour-coded in black. The total number of stellar particles inside
this sphere is approximately $27000$, out of which $23000$ belong to
the satellite and $4000$ to the host\footnote{This is due to higher
  number of particles used to simulate the satellite to properly
  resolve its structure in phase-space. Therefore, in practice only 1
  in 50 satellite particles should be considered.}.

The top panel of Fig.~\ref{fig:streams} shows the distribution of
particles in the $\Omega_{\phi}$ vs. $\Omega_{r}$ space. Note the
large amount of substructure present. As expected, satellite particles
are distributed in multiple lumps associated with the different
streams crossing this volume. These lumps are better defined for orbits
which spend most of their time far from the centre, and so are found at
low frequencies and have long mixing timescales.

The upper envelope defined by the smallest $\Omega_{r}$ at a given
$\Omega_{\phi}$ corresponds to particles with low eccentricity
(i.e. on more circular orbits) as discussed in Section
\ref{sec:char_space}. In our simulation this region is preferentially
populated by disk particles. 

In general, orbits in axisymmetric potentials have three independent
frequencies: the previously discussed $\Omega_{r}$ and $\Omega_{\phi}$
and a third frequency associated with the vertical motion,
$\Omega_{z}$. However, particles in the thick disk typically have very
short periods in the vertical direction. Hence, the number of streams
expected in the $\Omega_{z}$ direction is very large and therefore it
is hard to resolve in our simulation, and not discussed further here.

In the bottom left panel of Fig.~\ref{fig:streams} we can observe the
distribution of particles in the space defined by orbital apocentre
and pericentre. The stripy features along lines of different
eccentricities are the counterparts of the lumps located on diagonal
lines in frequency space. The middle panel shows the $E-L_{z}$ space,
where extended structures with slightly different orientations may be
seen. The panel on the right shows a projection of velocity space,
which clearly exhibits less well-defined structures in comparison to
the other spaces discussed. In general, we may conclude from this
Figure that streams are much better defined and easier to interpret in
frequency space than in any of the other projections considered.

\subsection{Estimating the time of accretion}
\label{sec:t_acc_nbody}
\begin{figure}
\centering
\includegraphics[width=63mm,clip]{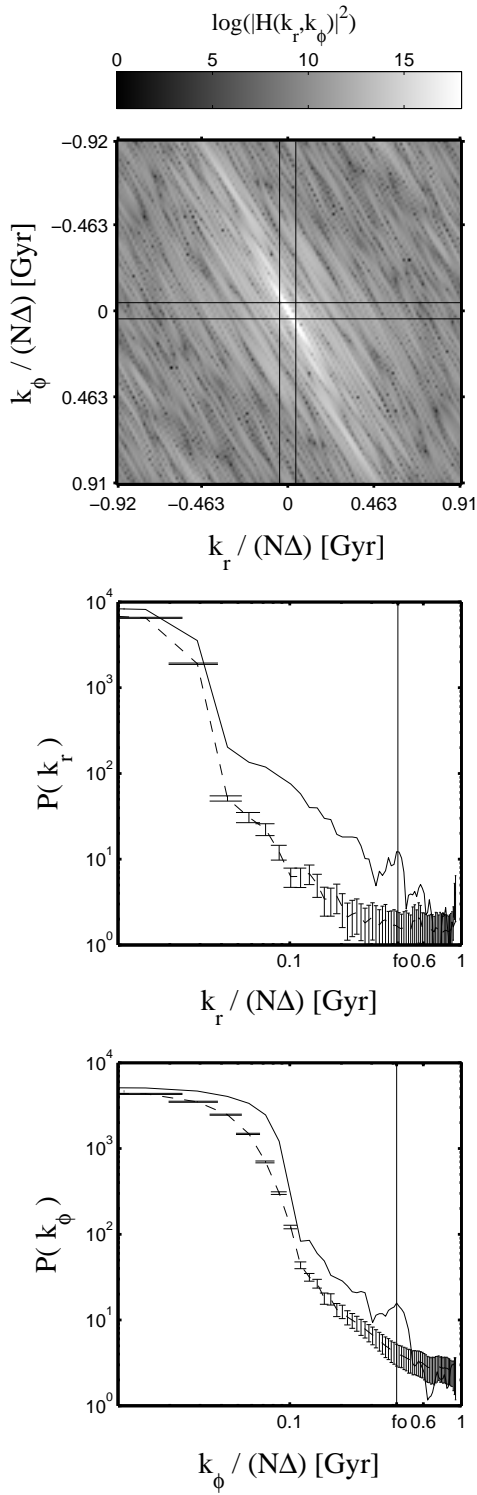}
\caption{Top panel: Fourier transform of the distribution of particles
  in frequency space shown in the top panel of
  Fig.~\ref{fig:streams}. As in Fig.~\ref{fig:ps_hr} the middle and
  central panels depict the 1-d power spectra around $k_\phi \sim 0$
  and $k_r \sim 0$, respectively. The vertical lines in these panels
  show the location of the peak associated to the average separation
  between streams in frequency space, and can be used to estimate the
  accretion epoch. The dashed curve corresponds to the average power
  spectrum obtained from a random distribution of particles in
  frequency space. The error bars represent the rms dispersion from
  this average spectrum.}
\label{fig:ps_thick_disk}
\end{figure} 

In comparison with the idealised cases discussed in Section
\ref{sec:comp} and \ref{sec:streams_t_evol} the distribution of
particles in frequency space for the VH08 simulation is less
regular. Nevertheless, we would like to explore if it is still
possible to obtain a good estimation of the time of accretion of the
satellite, using the Fourier analysis described in previous sections.

The top panel of Fig.~\ref{fig:ps_thick_disk} plots the image in
Fourier space of the frequency plane $\Omega_{\phi}$ vs. $\Omega_{r}$
discussed above. This image was obtained by making a grid in frequency
space and counting the number of particles in each grid element. As
before the black lines in the top panel of
Fig.~\ref{fig:ps_thick_disk} denote the cuts around $k_{r}=0$ and
$k_{\phi}=0$ made to compute the 1-dimensional power spectra.

To estimate the uncertainties in the power spectrum, we create a
second image obtained by assuming that the particles are spread
throughout the permitted regions of frequency space according to a
Poisson distribution. That is, we assign to each grid element $N_{p}$
particles, where this number is drawn from a Poisson distribution with
mean $\langle N \rangle = (N_{\rm disk} + N_{\rm sat})/n_{\rm grid}$,
i.e. equal to average number of stellar particles per grid element. We
then compute the Fourier transform of this distribution and the
corresponding 1-dimensional power spectra from the image. This
procedure is repeated 1000 times, to yield an average power spectrum
in each direction associated to a random distribution of particles in
frequency space.

The results are shown in the bottom and middle panel of
Fig.~\ref{fig:ps_thick_disk}. The black solid line in each panel
denotes the power spectrum obtained from the actual distribution of
particles in frequency space. The dashed line corresponds to the
average power spectrum for our random distributions where the vertical
error bars represent the rms dispersion around this average
spectrum. As we can see, a statistically significant peak can be
identified in the observed power spectrum along $k_r$ (and also for
$k_\phi$). The wavenumber of this peak is $f_{0}\approx 0.43$ Gyr for
both the radial and angular directions. This corresponds to an
estimated accretion time of $2.7$ Gyr. Recall that this simulation is
evolved for $4.5$ Gyr. However, the satellite only starts to lose a
significant amount of stars $\sim 1$ Gyr after infall (see Figure 3 of
VH08). Therefore, we may conclude that the time since disruption of a
satellite can be reasonably estimated even in a simulation with a live
host.

\section{Discussion and Conclusions}

We have confirmed and extended previous work by MB08 showing that orbital 
frequencies constitute a very suitable space to identify debris from past 
accretion events. In this space, particles in a given volume of 
physical space are found to populate well-defined lumps, 
each of them associated with a different stream.

For time-independent potentials, these streams are distributed in a
regular pattern along lines of constant frequency. As time goes by,
the number of streams at a given physical location increases while the
separation between adjacent streams decreases. We have shown here that
this characteristic separation or scale can be used to estimate the
time of accretion of the object through a Fourier analysis.

We have also addressed how the time evolution of the host affects
substructure in frequency space. As an example, we have considered the
case in which the mass of the host is increased exponentially in time
on two different timescales. We find that in contrast with the actions
which are adiabatic invariants for slowly varying potentials, the
frequencies always evolve in time, closely following the rate of
change of the host potential. This evolution however does not destroy
the clumpiness present in frequency space. Streams still look lumpy
and are regularly distributed in this space, even if the host
potential varies on a very short timescale (when even the actions are
no longer invariant). Interestingly in this case, streams are not
exactly distributed along straight lines of constant frequency but
rather on lines whose curvature depends on the timescale of growth of
the potential. This implies that, contrary to previous claims
\citep[e.g.][]{p06,wrck}, the final distribution of streams does
retain information on the evolution in time of the host.

Finally, we have analysed a full N-body simulation of the accretion of
a satellite galaxy onto a disk galaxy. Due to the inclusion of other
physical processes (such as self-gravity, dynamical friction and the
variation in time of the distribution of mass in the system), the
distribution of satellite particles in frequency space looks somewhat
less regular than in the previously discussed idealised
cases. Nevertheless, the space of frequencies is still rich in
substructure associated to streams. Furthermore, even in this case
Fourier analysis techniques can be used to estimate the time of
accretion reasonably well, although in general, only a lower limit is
obtained.

For all these examples we have compared the final distribution of
streams in the most commonly used spaces to identify satellite
debris. These comparisons have shown that frequency space contains
information that is either not present or is not simply obtained in 
other spaces. In general streams are most sharply defined in frequency
space.

One important simplification in our analysis is to consider the
accretion of a single satellite galaxy. In reality, the process of the
formation of a galaxy like the Milky Way will have involved many
mergers \citep[in the range of $5-40$, see][]{lh}. Therefore it is
possible, and even likely, that their debris will overlap in frequency
space. However, as we have shown, the separation between adjacent
streams in frequency space is predicted to be different for mergers
that took place at different epochs. Moreover, the location of their
debris in frequency space at the present time is dependent upon the
initial orbital conditions. Consequently we expect the overlap to be
incomplete and hence the identification of their remnants to be
feasible. We are currently analysing the full cosmological high
resolution N-body simulations of the Aquarius Project
\citep{springel2008} in order to address this issue and to understand
what to expect in the context of the hierarchical paradigm of
structure formation.

\section*{Acknowledgements}
We are very grateful to Alvaro Villalobos for providing the simulation
used in Section \ref{sec:nbody} and Daniel Carpintero for the software
for the spectral analysis used in Section \ref{sec:comp_freq}. FAG
 would like to thanks Saleem Zaroubi and Rajat Thomas for very 
helpful discussions. NWO is acknowledged for financial support through a
VIDI grant to AH.  We are indebted to the referee, Paul McMillan, for
a through, and insightful report that lead to improvements in the
presentation and content of this paper.

\label{lastpage}
\end{document}